\documentclass[11pt]{article}
\pdfoutput=1
\usepackage[utf8]{inputenc}
\usepackage[T1]{fontenc}
\usepackage{blois,amsmath,amssymb,lmodern,xcolor,graphicx,cancel}
\usepackage[merge,elide]{natbib}
\hypersetup{citecolor=blue}

\newcommand{\Photo}{}
\newcommand{\ind}[2]{\ensuremath{^{#1}_{\phantom{#1}#2}}}

\begin{document}
\begin{flushright}
CP3-13-46\vspace{-\baselineskip}
\end{flushright}
\vspace*{4cm}
\title{\uppercase{The same-sign top probe\\for baryon number violation\\at the LHC}\,\footnote{Talk given at the \emph{25th Rencontres de Blois}, May 2013, and
 based on works carried out in collaboration with J.-M. Gérard, F. Maltoni and C. Smith.\cite{Durieux:2012gj,Durieux:2013uqa}}}

\author{Gauthier Durieux}
\address{Centre for Cosmology, Particle Physics and Phenomenology (CP3), 
Université catholique de Louvain, 
Chemin du Cyclotron 2, 
B-1348 Louvain-la-Neuve, 
Belgium
}

\maketitle
\abstracts{
Despite strong experimental evidences for their conservation at low energies, the baryon $B$ and lepton $L$ numbers are not expected to be absolute symmetries. Already violated in the Standard Model (SM) by non-perturbative effects (tiny at low temperatures and energies) and not justified by any fundamental principle, $B$ and $L$ appear naturally violated in beyond the Standard Model (BSM) scenarios, \emph{e.g.} in supersymmetry (SUSY). The fact that any violation remained unobserved at low energies can nonetheless be explained, without pushing the associated characteristic sale beyond the TeV, by using SM flavour symmetries implemented BSM thanks to the Minimal Flavour Violation (MFV) prescription. Violation effects could therefore be observable at the LHC. Same-sign (anti-)tops, sometimes leading to a distinctive predominance of negatively charged lepton pairs over positively charged ones, is a clean signature that is for instance generically present in R-parity violating supersymmetry.}

\section*{Introduction}
\vspace*{-2mm}
The standard paradigm of the Minimal Supersymmetric extension of the Standard Model (MSSM) as a natural framework embedding a fundamental scalar at the electroweak scale is increasingly challenged by the constraints on sparticles masses extracted from LHC data. These limits however typically rely on the strong assumption of R-parity conservation that causes sparticles to be produced in pairs only and makes the lightest of them stable. A large amount of missing transverse energy is therefore usually imposed in SUSY searches.

The prime motivation for R-parity conservation arises from existing constraints most notably on (di)nucleon decays or neutron oscillation. Indeed, it forbids renormalizable interactions violating both $B$ and $L$,
\vspace*{-1mm}
\begin{equation*}
\lambda''_{abc}	\quad	\bar U^a\bar D^b\bar D^c,\qquad
\lambda'{}\ind{ab}{c}	\quad	Q_a L_b\bar D^c ,\qquad
\lambda\ind{ab}{c}	\quad	 L_a L_b\bar E^c,\qquad
\mu'{}^{a}	\quad	H_uL_a,
\vspace*{-1mm}
\end{equation*}
in principle allowed by MSSM symmetries. Order one $\lambda'',\lambda',\lambda,\mu'$ coefficients would otherwise require the SUSY scale to be raised to GUT energies. In view of the strong hierarchies observed in the SM Yukawa sector, it however looks unreasonable for these Yukawa-like coefficients to be all of the same order. As precise measurements of flavour transitions moreover tightly constrain flavour structures different from the SM ones, it is worth trying to estimate the hierarchies between RPV couplings using the knowledge of low-energy phenomena experimentally gathered over time and encoded in SM flavour structures.\cite{Nikolidakis:2007fc}

A consistent prescription to achieve this goal is provided by Minimal Flavour Violation. In this framework, BSM flavoured couplings are first formally made invariant under $G_F$, the global flavour symmetries of the SM gauge sector, by appropriate insertions of Yukawas that are treated as spurious fields (or \emph{spurions}) transforming non-trivially under $G_F$ (so as to make the Yukawa Lagrangian formally $G_F$-invariant too). They are then frozen to their physical values expressed in terms of fermion masses and CKM matrix elements. In this way are the BSM flavour structures consistently \emph{aligned} with SM ones.

As several spurion insertions can possibly make a given interaction $G_F$-invariant, the highest numerical value obtained after freezing is considered to provide an upper bound for the corresponding flavoured coupling. Without any further overall rescaling it was noted\cite{Nikolidakis:2007fc, Smith:2011rp, Csaki:2011ge} that on top of making BSM couplings consistent with flavour observables for a moderate characteristic scale, MFV was also actually making RPV interactions compatible with constraints on $B$ and $L$ violations for a SUSY scale of the order of the TeV.

\vspace*{-2mm}
\section*{Flavourful R-parity violation}
\vspace*{-2mm}
MFV imposes fairly strong constraints on the very structure of $B$ and $L$ violating interactions.

The flavour symmetries of the SM gauge sector are unitary rotations in generation space. $G_F$ is therefore a product of $SU(3_\text{g})$ factors (where $3_\text{g}$ is the number of generations), one for each kind of fermion field:
\vspace*{-1mm}
\begin{equation*}
G_F= SU(3_\text{g})_q
\times SU(3_\text{g})_l
\times SU(3_\text{g})_u
\times SU(3_\text{g})_d
\times SU(3_\text{g})_e
\vspace*{-1mm}
\end{equation*}
with $q$ and $l$, the quark and lepton doublets and $u,d,e$ the singlets. Each $SU(3_\text{g})$ factor only possesses two elementary invariant tensors: $\delta^a_b$ and $\epsilon_{abc}$. The $\delta^a_b$ tensor allows to build flavour invariant combinations of a flavoured field (transforming as a triplet) and its anti-particle (an anti-triplet). Such combinations, that are found in the SM gauge Lagrangian, however obviously conserves $B$ and $L$. The $\epsilon_{abc}$ tensor and appropriate spurion insertions can be used to construct $G_F$-invariant combinations of three flavoured fields. As SM Yukawas do not mix lepton and quark flavours, the only $G_F$-invariant and $B$ or $L$ violating combinations involve three flavoured fields transforming either as (anti-)quarks or (anti-)leptons.

Only the baryon number violating RPV interaction $\lambda''_{abc}\bar U^a\bar D^b\bar D^c$ satisfies this requirement. Though, extending the set of SM Yukawas so as to provide neutrinos with a Majorana mass (Dirac Yukawas do not help) make it possible to render $L$ violating RPV couplings formally $G_F$-invariant too. Nevertheless, they are then highly suppressed by a neutrino mass,\cite{Nikolidakis:2007fc, Smith:2011rp} making $|\Delta B|=1=|\Delta L|$ processes like proton decay slow.

Considering all possible Yukawa spurion insertions and retaining the combination leading to the mildest suppression for each of the $\lambda''$ couplings, those end up highly hierarchical.\cite{Nikolidakis:2007fc} For $\tan\beta=5$ and without overall rescaling, a value of about $0.1$ is obtained in this way for $\lambda''_{tds}$ while all other $\lambda''$ couplings suffer suppressions at least of the $10^{-4}$ order. Dinucleon decays and neutron oscillation that would primarily proceed through the couplings involving first generations only are therefore significantly suppressed.\cite{Nikolidakis:2007fc, Smith:2011rp, Csaki:2011ge}

\vspace*{-2mm}
\section*{LHC phenomenology}
\vspace*{-2mm}
Assuming that $\lambda''_{tds}$ dominates over the other couplings with a value of the order of $0.1$ and that the SUSY scale is nonetheless allowed to be of the order of the TeV, a significant production of same-sign top pairs is expected quite generically at the LHC. A detailed discussion was presented elsewhere\cite{Durieux:2013uqa} but a simplistic reasoning can be put forward to understand heuristically this conclusion:
\begin{list}{}{\itemsep=-1mm \leftmargin=5mm \topsep=2mm}
\item[1.] The (coloured) sparticles lying at the bottom of the SUSY spectrum are copiously produced.
\item[2.] Their RPV decay are often favoured kinematically.
\item[3.] As $\lambda''_{tds}$ dominates RPV couplings, tops or stops will often be involved in these decays.
\item[4.] Majorana gluino (and neutralino) make same-sign final state fairly common.
\end{list}
So we get many same-sign top pairs quite independently of the precise MSSM spectrum and parameters. This is a clear signature that can be identified in its clean same-sign dilepton and ($b$-)jets channel which therefore constitutes a discovery channel for MFV RPV in a wide variety of its variants (including the holomorphic MFV restriction\cite{Csaki:2011ge}).

One notable exception to this conclusion arises when a stop is much lighter than all other sparticles, including the gluino (otherwise $gg\to\tilde g\tilde g\to\bar t\tilde t\;\bar t\tilde t$ can be significant). After QCD pair production, each stop would then quickly decay into two light jets $\tilde t \to \bar s\bar d$ and, in such a scenario, dijet resonances are probably more promising signatures.

\vspace*{-2mm}
\section*{Representative simplified framework}
\vspace*{-2mm}
Two processes leading to same-sign tops are especially relevant at the LHC: $dd\to\tilde d_R\tilde d_R\to \bar s\bar t\;\bar s\bar t$ (or its conjugate)\quad and \quad $gg\to \tilde g\tilde g\to \bar t\bar d\bar s\;\bar t\bar d\bar s$ or $tds\;tds$. They respectively proceed through resonant sdown and gluino that are copiously produced when their masses lie around the TeV. The same-sign sdowns are produced through a t-channel gluino exchange while the gluino decays to $tds$ or $\bar t\bar d\bar s$ through a right down, strange or top squark (possibly on-shell).

\begin{figure}
\vspace*{-5mm}
\includegraphics[width=.5\textwidth]{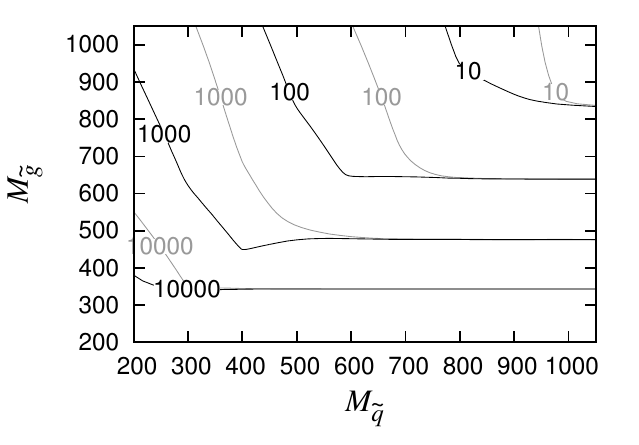}%
\includegraphics[width=.5\textwidth]{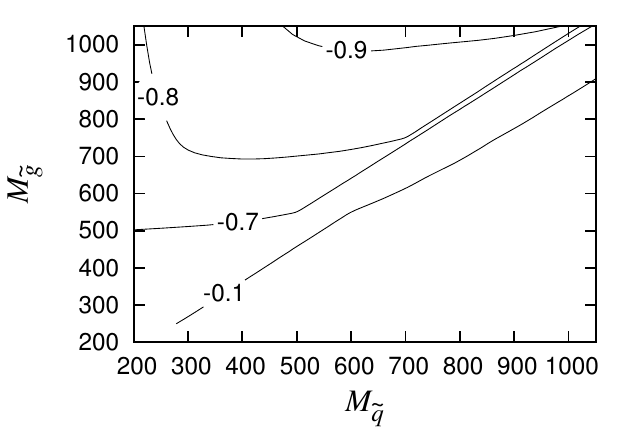}%
\vspace*{-5mm}
\caption{\emph{Left.} Cross section [fb] at the $8$~TeV LHC for same-sign (anti-)top production through resonant sdown and gluino productions followed by RPV decays via the $\lambda''_{tds}=0.1$ coupling. The gluino and a common quark mass are varied while all other sparticles are decoupled. Those leading order and parton level results have been obtained using {\sc FeynRules}\cite{Christensen:2008py} and {\sc MadGraph5}.\cite{Alwall:2011uj} The black contours are obtained when the mostly-right sdown only contributes to the signal and the grey ones when the mostly-left sdown contributes as much as the mostly-right one. \emph{Right.} Corresponding asymmetry between top and anti-top pair production rates: $\sigma_{tt}-\sigma_{\bar t\bar t}\big/\sigma_{tt}+\sigma_{\bar t\bar t}$.}
\vspace*{-5mm}
\label{fig1}
\end{figure}

In the limiting case where all sparticles but the sdown and gluino decouple, with one of these two sparticles significantly lighter than the other, its RPV branching fraction will tend to $100\%$. The same-sign top rate (see Fig.~\ref{fig1}) then only depends on the $M_{\tilde d}$ and $M_{\tilde g}$ masses.

A very striking feature of the $dd$-initiated process is that it leads to a significant predominance of same-sign anti-tops over tops (see Fig.~\ref{fig1}). This fact, caused by the higher parton distribution functions of valence quarks with respect to their anti-particles, leads to a charge asymmetry in the same-sign dilepton production rate: more negative than positive leptons are produced by this process. In a proton collider, this distinctive feature is very much characteristic of baryon number violation.\cite{Durieux:2012gj}

\vspace*{-2mm}
\section*{Approximate limits}
\vspace*{-2mm}
The CMS collaboration has performed a same-sign plus ($b$-)jets search in $10.5$ inverse femtobarns of $8$~TeV data.\cite{Chatrchyan:2012paa} Amongst the signal regions defined, the one characterised by two same-sign isolated leptons ({\small $p_T>20$~GeV, $|\eta|<2.4$}), four jets ({\small $p_T>40$~GeV, $|\eta|<2.4$}), two b-tags, a high hadronic activity $H_T>320$~GeV and no missing transverse energy $\cancel{E}_T$ requirement currently provides the best sensitivity to the same-sign tops RPV signal. In the future, if the limit is raised to higher sparticles masses, the signal missing transverse energy carried away by neutrinos may increase enough for a $\cancel{E}_T$ cut, suppressing a lot of background, to be advantageous.

We used a simplified analysis mimicking the CMS one to derive approximate limits in the $M_{\tilde g}-M_{\tilde q}$ plane. Most notably, events were generated with {\sc MadGraph5}\cite{Alwall:2011uj} at leading order and parton level, and $b$-tagging as well as isolated lepton selection efficiencies were fixed at $60\%$ (in agreement with what the $p_T$ parametrisation provided by CMS\cite{Chatrchyan:2012paa} gives for the signal considered).

\begin{figure}
\vspace*{-5mm}
\includegraphics[width=.5\textwidth]{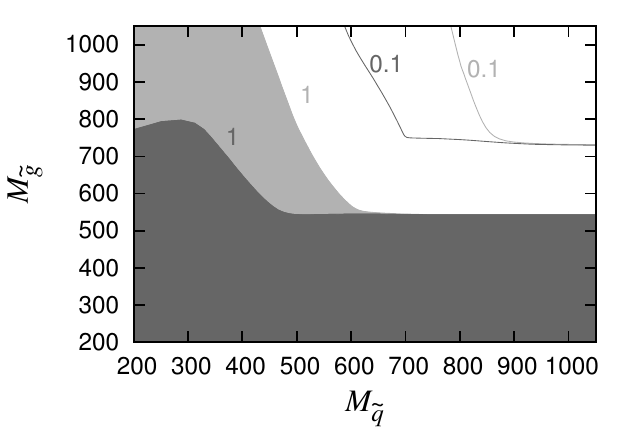}
\includegraphics[width=.5\textwidth]{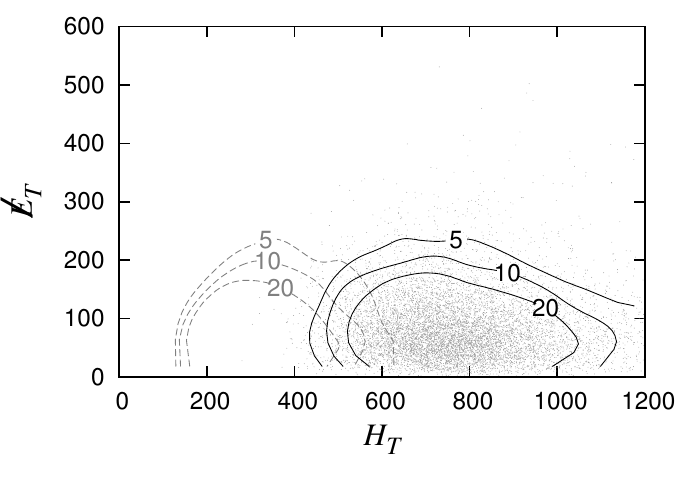}
\vspace*{-1cm}
\caption{\emph{Left.} Fiducial cross section at the $8$~TeV LHC for same-sign dilepton plus ($b$-)jets production with $H_T>320$~GeV and $\cancel{E}_T>0$~GeV, as a function of the squarks and gluino masses. The dark grey region is excluded by the corresponding $10.5$/fb CMS search\cite{Chatrchyan:2012paa} when only the mostly-right sdown contributes to the signal and the limit excludes the light grey region too when the mostly-left sdown contributes as much as the mostly-right one.
\emph{Right.} Signal $H_T-\cancel{E}_T$ shape for $M_{\tilde g}=550$~GeV, $M_{\tilde q}=700$~GeV (black) compared to the $t\bar t W+t\bar t Z$ SM background (dashed grey). Numerical values for $1/\sigma\;\text{d}^2\sigma\big/\text{d}H_T\text{d}\cancel{E}_T$ are given in units of $(100~\text{GeV})^{-2}$.}
\vspace*{-5mm}
\label{fig2}
\end{figure}

In the $M_{\tilde q} < M_{\tilde g}$ region, the limit (see Fig.~\ref{fig2}) significantly depends on the exact MSSM parameters (\emph{e.g.}, whether the mostly-left sdown contributes as much as the mostly-right one or not at all). The $M_{\tilde g}$ dependence of the same-sign sdown cross section is manifest and a loss of efficiency for the lowest $M_{\tilde q}$'s considered can be attributed to the high hadronic activity and jet multiplicity requirement. In the $M_{\tilde g}<M_{\tilde q}$ region, the limit obtained is fairly robust and approximately constrains the gluino to be no lighter than $550$~GeV.

\vspace*{-2mm}
\section*{Conclusions}
\vspace*{-2mm}
LHC constraints on supersymmetric particle masses, typically derived assuming a large missing transverse energy, are dramatically rising and challenging naturalness in SUSY scenarios. The paradigm of R-parity conservation that motivated such a missing energy requirement is however questionable. This \emph{ad hoc} assumption can however be relaxed without introducing unnaturally massive superparticles if what we know about flavour transitions in the Standard Model is extrapolated beyond it, using the Minimal Flavour Violation prescription. Same-sign tops are then generically produced and clearly identified in their same-sign leptons and ($b$-)jets channel. A predominance of negatively charge dileptons, very distinctive at the LHC, is present when valence quark initiated processes contribute to that signal.

\vspace*{-2mm}
\section*{Acknowledgements}
\vspace*{-2mm}
J.-M. Gérard, F. Maltoni and C. Smith are warmly thanked here for the enjoyable collaborations that led to the presented results. The author is a Research Fellow of the F.R.S.-FNRS, Belgium.

\begin{flushleft}
\setlength{\bibsep}{.4mm}
\vspace{-2mm}
\bibliographystyle{apsrev4-1-moriond}
\bibliography{blois.bib}
\end{flushleft}
\end{document}